\documentclass[12pt,preprint]{aastex}

\newcommand{\gsim}{\mbox{\hspace{.2em}\raisebox{.5ex}{$>$}\hspace{-.8em}\raisebox{-.5ex}{$\sim$}\hspace{.2em}}}
\newcommand{\lsim}{\mbox{\hspace{.2em}\raisebox{.5ex}{$<$}\hspace{-.8em}\raisebox{-.5ex}{$\sim$}\hspace{.2em}}}
\newcommand{\ssst}{\scriptscriptstyle}
\newcommand{\E}[1]{\times 10^{#1}}
\newcommand{\etal}{et al.}
       
\newcommand{\RA}[3]{{#1}^{{\rm h}}{#2}^{{\rm m}}{#3}^{{\rm s}}}
\newcommand{\Dec}[3]{{#1}^{\circ}{#2}'{#3}''}

\newcommand{\s}{\,{\rm s}}      \newcommand{\ps}{\,{\rm s}^{-1}}
\newcommand{\yr}{\,{\rm yr}}    \newcommand{\Msun}{M_{\odot}}
\newcommand{\cm}{\,{\rm cm}}    \newcommand{\km}{\,{\rm km}}
\newcommand{\parsec}{\,{\rm pc}}\newcommand{\kpc}{\,{\rm kpc}}
\newcommand{\ergs}{\,{\rm ergs}}        \newcommand{\K}{\,{\rm K}}
    \newcommand{\keV}{\,{\rm keV}}

\newcommand{\nel}{n_{e}}        \newcommand{\NH}{N_{\ssst\rm H}}
\newcommand{\no}{n_{\ssst 0}}   
 
\newcommand{\Ts}{T_{s}}
\newcommand{\rs}{r_{s}}         \newcommand{\vs}{v_{s}}
\newcommand{\nH}{n_{\ssst\rm H}}        \newcommand{\mH}{m_{\ssst\rm H}}
\newcommand{\nHH}{n({\rm H}_{2})} 
\newcommand{\xray}{X-ray}       \newcommand{\Einstein}{{\sl Einstein}}
\newcommand{\ROSAT}{{\sl ROSAT}} \newcommand{\ASCA}{{\sl ASCA}}
\newcommand{\Chandra}{{\sl Chandra}}
\newcommand{\du}{d_{8}}         \newcommand{\Eu}{E_{51}}
   \newcommand{\ru}{r_{3}}

\begin{document}

\title{A \Chandra\ ACIS view of the Thermal Composite Supernova Remnant 3C~391}
%\title{Highly Clumpy Structure of the Thermal Composite Supernova Remnant 3C~391 Unveiled by \Chandra\ Observations}

\author{
 Yang Chen\altaffilmark{1,2},
 Yang Su\altaffilmark{1},
 Patrick O.\ Slane\altaffilmark{3}, and
 Q.\ Daniel Wang\altaffilmark{2}
}
\altaffiltext{1}{Department of Astronomy, Nanjing University, Nanjing 210093,
       P.R.China}
\altaffiltext{2}{Department of Astronomy, B619E-LGRT, 
       University of Massachusetts, Amherst, MA01003}
\altaffiltext{3}{Harvard-Smithsonian Center for Astrophysics,
Cambridge, MA 02138}

\vfil
%Text
\begin{abstract}
We present a 60 ks \Chandra\ ACIS-S observation of the thermal
composite supernova remnant 3C~391.
The southeast-northwest elongated morphology is similar to that
previously found in radio and X-ray studies.
This observation unveils a highly clumpy structure of the remnant.
Detailed spatially resolved spectral analysis for the small-scale
features reveals that the interior gas is generally of normal
metal abundance and has approached or basically reached 
ionization equilibrium.
The hydrogen column density increases from southeast to northwest.
Three mechanisms, radiative rim, thermal conduction,
and cloudlet evaporation, may all play roles in the X-ray appearance
of 3C~391 as a ``thermal composite" remnant,
but there are difficulties with each of them in explaining
some physical properties.
Comparatively, the cloudlet evaporation model is favored by
the main characteristics such as the highly clumpy structure and
the uniform temperature and density distribution over most of the remnant.
%and the much lower X-ray emitting gas density than the mean ambient
%cloud density.
%strongly favors the cloudlet evaporation model, one of the
%candidate mechanisms presently available to account for
%the X-ray appearance of 3C~391.%as a ``thermal composite" remnant. 
The directly measured postshock temperature also implies a young
age, $\sim4\E{3}\yr$, for the remnant.
The postshock gas pressure derived from the NE and SW rims, which harbor
maser spots, is consistent with the estimate for the maser regions.
An unresolved X-ray source is observed on the northwest border
and its spectrum is best fitted by a power-law.
%Its spectrum is best fitted by a power-law, and in view of its position and
%absorption, the possibility of the association
%with 3C~391 cannot be ruled out.
% make it a possible candidate for a compact object associated with 3C~391.


\end{abstract}

\section{Introduction}
3C 391 (G31.9+0.0), a supernova remnant (SNR) with irregular
morphology, has been observed in several electromagnetic bands.
Radio observations with the VLA reveal an elongated structure
extending from the
northwest (NW) to the southeast (SE), surrounded by a shell
except on the SE border, where it appears that the SNR has broken
out of a dense region into an adjacent region of lower density
(Reynolds \& Moffett 1993).
In X-rays, both \Einstein\ (Wang \& Seward 1984)
and \ROSAT\ (Rho \& Petre 1996) observations show that
the centroid of the soft X-ray emission sits in the SE region.
The \ASCA\ observation reveals a hard X-ray enhancement
in the NW region, and confirms the decrease in hydrogen column
density across the remnant from NW to SE (Chen \& Slane 2001).
Two hydroxyl radical 1720 MHz maser spots are found along the radio
shell (Frail et al.\ 1996).
The masers and the strong enhancement in [OI] $63\mu$m
emission near the northwestern edge (Reach \& Rho 1996) both
provide evidence for shock interactions with a molecular cloud.
CO and other molecular line
observations place the location of the remnant at the
southwestern (SW) edge of a molecular cloud
(Wilner, Reynolds, \& Moffett 1998; Reach \& Rho 1999).

%Theoretical significance, cloud evaporation for instance
3C~391 is an example of the ``thermal composite'' or ``mixed
morphology'' category of supernova remnants, which
also includes W28, W44, IC~443, G349.7+0.2, and others
(Rho \& Petre 1998).
They generate bright {\em thermal} X-ray emission interior to
their radio shells, and have faint X-ray rims.
They are usually found to interact with
adjacent molecular clouds, characterized by the hydroxyl radical
maser emission (Green \etal\ 1997, Yusef-Zadeh et al.\ 2003).
Non-thermal diffuse X-ray emission inside SNRs is widely believed
to be synchrotron radiation from pulsar wind nebulae,
while the nature of the internal thermal X-ray emission
seen in mixed morphology remnants
is still uncertain.
%So far three candidate scenarios compete to account for the mechanism
%of the internally-brightened X-ray morphology.

At least four distinct scenarios have been proposed to explain
centrally-brightened X-ray morphology.
%or by an SNR which has entered the radiative stage with the interior
%gas still hot, but with the rim material cooled down.
%Since the gas at the rim is usually denser and colder than the gas
%in the inner volume, 
The first scenario is radiative cooling of the rim gas.
Under this hypothesis, the gas at the rim has been cooled down in the
radiative stage of evolution, with a temperature so low ($<10^6\K$) that its
X-ray emission is very weak, while the gas in the inner volume
is still hot enough to emit strong X-rays (e.g.\ Harrus et al. 1997,
Rho \& Petre 1998).%, Cox et al.\ 1999, Shelton et al.\ 1999).
The second mechanism invokes thermal conduction.
%It is suggested that, in its young age, thermal conduction
%in the remnant can prevent formation of the very tenuous, hot gas
%in the inner part that is characteristic of the adiabatic expansion
%and therefore change the interior structure from the standard
%Sedov solution, resulting in a nonnegligible density and luminous
%X-ray in the interior (as observed in, especially, the radiative stage)
It is suggested that thermal conduction in the remnant can smooth
out the temperature gradient from the hot interior to the cooler shell,
and increase the central density in response to the associated
change in pressure
conditions. This results in luminous X-rays in the interior, and is
dominant in the radiative stage
(Cox et al.\ 1999, Shelton et al.\ 1999).
The third scenario invokes cloudlet evaporation in the SNR interior.
When a SNR expands in an inhomogeneous interstellar medium (ISM)
whose mass is mostly contained in small clouds,
the clouds engulfed by the blast wave can be evaporated
to slowly increase the density of the interior gas;
as a result, the SNR appears internally X-ray brightened
(White \& Long 1991).
The fourth suggestion is that the mixed morphology is a projection effect.
For shell-like SNRs
that evolve in a density gradient such as at the edge of a molecular
cloud, if the line of sight is essentially aligned with the density
gradient as well as the magnetic field, the SNRs will appear as
thermal composites (Petruk 2001).

To date, X-ray studies have not yet provided strong evidence
favoring any one of these mechanisms. %and
Here we report on high resolution \Chandra\ observations of 3C~391
aimed at addressing the nature of the centrally filled X-ray
emission in this SNR.
%the physics of the thermal composite SNRs is still uncertain,
%The \Chandra\ observation, with its unprecedented resolution,
%facilitates X-ray spectro-spatial analysis of the physical
%properties of 3C~391 and would help ascertain the nature
%of the internal thermal X-ray emission.

\section{Observation and Data Analysis}
We observed SNR 3C~391 with the Advanced CCD Imaging Spectrometer
(ACIS) on board the \Chandra\ X-ray observatory
on 03 August 2002 (Obs.\ ID 2786) for an exposure of 62 ks.
% target (282.367917, -0.9375)
The target center ($\RA{18}{49}{28}.3$, $\Dec{-00}{56}{15}$) was placed
$\sim 2'$ off the telescope aim-point for an optimal coverage of
the supernova remnant on the S3 (\#7) CCD chip (using very faint mode).
%with an Y-offset of $-2'$ from the nominal aim-point.

The level 1 raw event data were reprocessed to generate a new level 2
event file to capitalize on the Chandra Interactive
Analysis of Observations (CIAO) software package (version 2.3).
In the reprocessing, bad grades have been filtered out and good time
intervals have been reserved.
We also removed time intervals with significant background flares
(peak count rates $\gsim 3\sigma$ and/or a factor of $\gsim 1.2$ off
the mean background level of the observation).
% using Maxim Markevitch's
%light-curve cleaning routine ``lc\_clean''
%\footnote{available at http://hea-www.harvard.edu/~maxim/axaf/acisbg/}.
This cleaning, together with a correction for the dead time of the
observation, resulted a net 60.7 ks exposure for subsequent analysis.
Furthermore, a correction was made for the time-dependence of the
ACIS gain.

\subsection{Spatial Analysis}\label{sec:img}
For imaging analysis, we produced exposure maps in
0.3-1.5, 1.5-3.0, and 3.0-7.0 keV bands.
These maps were used for flat-fielding, accounting for bad pixel removal
as well as correcting for telescope vignetting
and the quantum efficiency variation across the detector.
The images in the three bands are adaptively smoothed
(using CIAO tool {\em csmooth} with signal-to-noise ratio of 3) and
exposure-corrected.
The tri-color X-ray image (0.3-1.5 keV in red, 1.5-3.0 keV in green, and
3.0-7.0 keV in blue), overlaid with the 1.5~GHz radio contours
(from Moffett \& Reynolds 1994), is shown in Fig.1a.
Another tri-color image with 1.5~GHz emission in red, 0.3-3.0~keV
in green, and 3.0-7.0~keV in blue is shown in Fig.1b.
A broad band (0.3-7.0~keV) image overlaid with the radio contours
are shown in Fig.1c.
The two OH maser points (Frail et al.\ 1996) are labeled in
these three maps.

We searched for point-like sources in three broad bands,
0.3-1.5 keV (S), 1.5-7.0 keV (H), and 0.3-7.0 keV (B).
A combination of source detection algorithms were applied:
wavelet, sliding-box, and maximum likelihood centroid fitting
(see Wang, Chaves, \& Irwin 2003 for details of the approach).
The estimation of the count rate of a source was based on the data
within the 90\% energy-encircled radius (EER) determined with
the calibrated point spread function of the instrument.
The information obtained for the point-like sources detected
this way on the S3 chip are summarized in Table~1.

In order to produce a map of diffuse emission, we removed the
point-like sources with each source region as a circle of twice
the 90\% EER.
After filling the source-removed ``holes'' (in a swiss-cheese-like
map) by interpolating the intensity from the surrounding regions,
we produced a smoothed (with count-to-noise ratio of 6),
exposure-corrected diffuse X-ray map in 0.3-7 keV (Fig.1d).
Also shown in Fig.1d are the regions used for spectral fitting
(see \S2.2).

%Since a considerable contribution of the remnant's X-ray emission comes from
%the lines of metal species Mg, Si, and S (\S2.2),
%we also present the narrow band 1.2-1.5, 1.7-2.0, and 2.3-2.6$\keV$
%diffuse emission images (including Mg He$\alpha$, Si He$\alpha$,
%and S He$\alpha$, respectively) in Fig.2,
%which are {\em csmooth}ed with signal-to-noise
%ratio of 3 and exposure-corrected.
%%(without the correction for the underlying continuum contribution).

These X-ray images display the SE-NW elongated morphology,
similar to that observed in the earlier radio (Moffett,
\& Reynolds 1994) and X-ray (Rho \& Petre 1996, Chen \& Slane 2001)
observations, but with considerable small-scale structure
revealed by the superb angular resolution of \Chandra.

The X-ray maps reveal a highly clumpy structure of the remnant,
with clumps or knots located in both the SE and NW parts.
A bright hard enhancement is peaked at
($\RA{18}{49}{27}.0, \Dec{-00}{54}{10}$) near the the NW border
(see Fig.1a); this corresponds to region \#3 in Fig.1d.
A complex mixture of knots is seen in the SE part of the remnant interior,
including at least four bright enhancements in all colors centered at
$\sim(\RA{18}{49}{29}.3,\Dec{-00}{56}{42})$ (region \#6),
    ($\RA{18}{49}{32}.9$, $\Dec{-00}{56}{58})$ (region \#7),
    ($\RA{18}{49}{35}.5$, $\Dec{-00}{57}{04})$ (region \#8), and
    ($\RA{18}{49}{29}.8$, $\Dec{-00}{57}{45})$ (region \#9).
Similar clumpy structures are also seen in the interior of SNR~N49B
(Park et al.\ 2003).
%and SNR 0103$-$72.6 (Park et al.\ 2003b).
Several remarkable, very bright knotty features appear
on the east and SE border of 3C~391.
%These features seems to be dense, bullet-like gaseous knots
%that move fast and shock against the interstellar medium.
Also, the eastern knot seems to coincide with the location of the 
infrared source IRAS~18470-0100 (whose fluxes are 1.32, 4.66, 23.8,
and 236 Jy at 12, 25, 60, and 100 $\mu$m, respectively).
%We cannot judge whether this infrared source is associated with
%the remnant or not.
These bright knots on the border may be small clouds that have
recently been shocked by the supernova blast wave.
%Though, the eastern knot appears to have a sharp interface
%toward the exterior of the remnant and a trailing extension backward
%to the interior (Fig.1c),
%bearing a resemblance with the Vela SNR fragment A
%observed by {\sl XMM}-Newton EPIC (Aschenbach 2002).

In the RGB 3-color map Fig.1a, yellow and light-green colors indicate
strong soft emission in the SE part, while fainter soft emission
appears brown in color and is seen in the middle section
of the elongated body of the remnant.
A faint arc appears in the rim of the northern brown patch.
This arc/shell seems to connect with another shell like structure
at the northeastern (NE) rim, which is evident in radio (Fig.1b)
and discernible in X-rays,
and harbors one of the previously detected OH masers (Frail et al.\ 1996).
On the opposite side, no shell-like X-ray structure can be discerned
along the southern and SW border, but faint diffuse X-ray emitting
gas seems to spill outside the radio border in the southwest.

In Fig.1a, the green color representing the middle-band emitting gas
almost pervades the whole postshock region of the NW part.
The pervasive green color in the NW, mixed with a small blue patch,
shows hardening of the X-ray spectrum.
This is consistent with gradual increase in hydrogen column density
across the remnant (Chen \& Slane 2001; see also \S3.2 below).
%%(see the next section of {\sl spectral analysis}).
%molecular cloud resides,

The radio contour overlays make it very apparent that there is a
shock interaction with a molecular cloud on the NW border
because of the flat morphology of the radio emission.
At the west rim, a soft X-ray brightened slab appears to be
very close to (just slightly behind) the radio peak emission (region \#12),
and may be related to a small dense region there.

A few point sources are seen in the remnant's
geometric center and the south and west portions (Fig.1 and Table 1).
The small number of counts collected
make it difficult to determine their nature.
A bright hard enhancement ($\RA{18}{49}{22}.3, \Dec{-00}{53}{34}$)
appears as an unresolved source on the NW border
(source \#17 in Table 1; region \#5 in Fig.1d).
We discuss the spectrum of this source below (\S3.3.).

\subsection{Spectral Analysis}\label{sec:spec}
Taking advantage of \Chandra's superb angular resolution, we
have performed the spatially-resolved spectral analysis of 3C~391.
Prior to the extraction of spectra, we removed the point-like
sources detected on the S3 chip (see Table 1) except for weak
sources with $\log(P)>-9.0$ (see Table 1 for the detailed definition 
of the false detection probability $P$)
and those with a high possibility of being clumps of
diffuse emission (judged by eye).
We defined 13 small-scale regions, 3 medium-scale regions, and the
region of the entire remnant (as diagramed in Fig.1d) for spectral investigation.
Most of the small-scale regions are chosen to include the small features
of X-ray enhancement such as the knots and the faint shell like structures. 
Two medium-size circles are used to compare the mean spectral variation
between the inner SE and NW portions. 
The remaining area on the S3 chip outside region \#17 for the entire remnant
was used for background.
For each chosen region, we use the CIAO script {\em acisspec} to extract the
spectrum and obtain the weighted response matrix.
%(which incorporates the ACIS response matrices generated by
% Townsley, L.K., Broos, P.S, Chartas, G.\ et al.\ 2002, Nuclear
% Instrument \& Methods in Physics Research Section A, 486, 716)
The net count rates of these extracted spectra are tabulated in
Table~2 and Table~3.
Corrections for the continuous degradation in the ACIS quantum efficiency
have been applied to auxiliary response files.
All the spectra mentioned above were regrouped to contain at least
25 net counts per bin.

There are three distinct line features,
Mg~He$\alpha$ ($\sim1.35\keV$),
Si~He$\alpha$ ($\sim1.85\keV$), and
S~He$\alpha$ ($\sim2.46\keV$)
in these spectra except the
spectrum of a NW unresolved source (region \#5),
confirming their thermal origin (see Fig.\ 2 and Fig.\ 3).
%The Fe L complex (at $\sim0.5$-$1.5\keV$) could have significant
%contribution in the spectra though it could not be constrainted well.
The Fe L complex could have significant
contribution in the range $\sim0.5$-$1.5\keV$.
The emission diminishes rapidly above the photon energy $\sim 5\keV$
and no Fe~K$\alpha$ emission is observed.

Using the Morrison \& McCammon (1983) interstellar absorption, we
test to fit the thermal spectra with various collisional equilibrium
ionization and non-equilibrium ionization (NEI) models
for the optically thin plasma of SNRs available in the XSPEC code.
We find that our spectra can be best described by the VNEI
model with NEI spectral version 2.0, which uses the Astrophysical
Plasma Emission Database (APED) to calculate spectrum
\footnote{http://cxc.harvard.edu/atomdb/sources.html}.
Considering the evident line emission from Mg, Si, S, and possibly
Fe L, we treat the abundances of these element species as
free parameters.
In most cases, however, thawing the abundances did not significantly
improve the spectral fit relative to that using solar abundances.
For a few cases, especially for the emission from medium-scale
regions and from the entire remnant, thawing the abundances do
indeed improve the fitting.
The spectral fit results are tabulated in Table~2.
The spectrum of the NW unresolved source (region \#5) can be best fitted
with an absorbed power-law emission (see the spectral fit parameters
summarized in Table~3) and cannot be fitted with a blackbody emission.
%Models of thermal bremsstrahlung and blackbody were also tried
%but yielded worse fits to the data than power-law.

The spectral fits show that the diffuse emission from various regions
have ionization parameters ($n_e t$) close to or higher than
$10^{12}\cm^{-3}\s$.
This implies that the hot plasma in the SNR is very close to, or is
basically in, the ionization equilibrium.

The spectral fits also show that the diffuse emission from small-scale
regions can be well fitted with solar abundances, and the emissions
from larger regions can be fitted with abundances very close to solar
values.
%The spectral fit for the knots on the SE border (region \#1)
%presents a hint of enriched metals with slightly elevated Mg, Si, S, and
%Fe abundances.% (if these abundances are free parameters).

The temperature of the gas interior to the SNR is% found generally around
$\sim$0.5-$0.6\keV$, with only small fluctuations.
For instance, the lowest temperature ($\sim0.46\keV$) is found at the
faint SE rim (region \#13), and the highest temperature ($\sim0.8\keV$)
%is seen at the bright NW knot (region \#4) and the NE rim (region \#10,
is seen at the NE rim (region \#10).
%shock front).

The absorption column density is found to generally increase across
the remnant from SE to NW.
This is consistent with the conclusion of the \ROSAT\ (Rho \& Petre 1996)
and \ASCA\ studies (Slane \& Chen 2001),
and is also consistent with the emission hardening shown in the previous
section.
The absorption in the SE part is about $\sim2.9\E{22}\cm^{-2}$ (region \#14).
The highest absorption ($\sim4.1\E{22}\cm^{-2}\ps$) is found in a NW
region (region \#4) which is in blue in the tri-color image (Fig.1a).
This hardest region was also seen in the \ASCA\ study
(see Fig.3 in Chen \& Slane 2001).
A very high absorption ($\sim3.7\E{22}\cm^{-2}\ps$) is also found at
the westmost rim (region \#12).
The difference of mean $\NH$ between the central SE region (region \#14)
and the central NW region (region \#15) is $\sim6\E{21}\cm^{-2}$.
Such a variation in $\NH$ is consistent with the existence of
a molecular cloud to the NW (Wilner et al.\ 1998;
Reach \& Rho 1999).

\section{Discussion}
%\subsection{Some parameters}
Similar to the hydrogen column density distribution across the remnant
obtained in previous X-ray studies, the $\NH$ values obtained in the
spectral analysis are confirmed to increase from the SE toward the NW.
The spatial analysis also shows an obvious spectral hardening in
the NW part.
This could naturally be explained by the scenario that the NW part
of the remnant is embedded in the molecular cloud
(Reynolds \& Moffett 1993).
The mean difference of $\NH$ between the SE and NW interior
portions, $\sim6\E{21}\cm^{-2}$, is consistent with that
found in the \ASCA\ study,
and implies a mean density
of the ambient molecular gas
$\langle\nHH\rangle$ of order $\sim10$-$20\cm^{-3}$
on the assumption that the SNR line-of-sight depth in the cloud
is similar to the remnant radius (Chen \& Slane 2001).
%based on the \ASCA\ study;
We also adopt the distance $d\sim8\du\kpc$ estimated there.

%We note that, if the remnant were embedded in the cloud,  
%The reason why the temperatures are similar between the NW
%and SE parts is indeed unclear, and thermal conduction should play a role
%in this regard. There may be other hydrodynamic factors, for example, the
%blowout into the low-density region may decrease the interior pressure
%(and the energy density) of the buried NW part, and hence lower the
%anticorrelation between the temperature and medium density.

According to the NEI fit to the spectrum of the entire remnant
(region \#17), the X-ray luminosity in 0.5--10 keV of the remnant
%is $L_X=2.3\E{36}\du^{2}\ergs\ps$.
is $L_X=3.5\E{36}\du^{2}\ergs\ps$.
%From the volume emission measure $f\nel\nH V \sim1.4\E{59}\du^{2}\cm^{-3}$,
From the volume emission measure $f\nel\nH V \sim1.5\E{59}\du^{2}\cm^{-3}$,
where $f$ is the filling factor of the \xray\ emitting gas,
the mean number density of the gas interior to the remnant is
%$1.8f^{-1/2}\du^{-1/2}\cm^{-3}$ (here $\nel\approx1.2\nH$ has been used
$1.9f^{-1/2}\du^{-1/2}\cm^{-3}$ (here $\nel\approx1.2\nH$ has been used
and a cylinder-like volume of a diameter $5'$ and a height $7'$
has been assumed).
The remnant volume is $V\sim5.1\E{58}\du^3\cm^{-3}$ .
The mass of the X-ray emitting gas is
%$M_{x}\sim1.4\nH\mH fV\sim108f^{1/2}\du^{5/2}\Msun$.
$M_{x}\sim1.4\nH\mH fV\sim114f^{1/2}\du^{5/2}\Msun$.

Raw estimates for the gas density of the regions used for spectral fit
(apart from \#17, the entire remnant)
can also be given in terms of the volume emission measures.
We assume that the circular regions correspond
to spheres, the elliptical regions to oblate spheroids, and the
rectangular regions to shell structures.
The gas densities thus obtained are listed in the last column of Table 2.
The densest X-ray emitting gas is located in the SE and eastern
regions \#1 and \#2 that contain very bright knots
(with density $\gsim10f^{-1/2}\du^{-1/2}\cm^{-3}$).
The other bright knots (regions \#3 in the NW and \#6, \#7, \#8, and \#9
in the inner central SE portion) have a gas density 
$\sim5$--$7f^{-1/2}\du^{-1/2}\cm^{-3}$.
The regions along the remnant border (\#10, \#11, \#12, \#13, and \#16)
have a density $\sim1$--$3f^{-1/2}\du^{-1/2}\cm^{-3}$.

%knots (SE, E, N, C1, C2, C3, SC)
%each with X-ray flux of order $\sim10^{-11}\ergs\cm^{-2}\ps$)\\
%C3: $\nH\sim5.4f^{-1/2}\du^{-1/2}$, $\sim0.45f^{1/2}\du^{5/2}\Msun$
%causing it to break-up and fragment.

%\subsection{Mechanism of internal thermal emission and
%dynamics of the remnant}
\subsection{Origin of centrally-filled X-ray morphology and
dynamics of the remnant}

As mentioned in the Introduction, four main mechanisms
have been proposed to explain the X-ray morphology of thermal
composite SNRs: projection effect,
radiative shells, thermal conduction, and cloudlet evaporation.
Here we compare them with the properties found from our
spatially-resolved spectral analysis.

\subsubsection{Projection effect}

The projection effect does not match the properties 
found in 3C~391.
As mentioned above, the molecular cloud is located in the NW but
the X-ray emission is not only enhanced in the NW half, but also
in the SE half.
Moreover, the hydrogen column density increases along the SE-NW elongation
direction, so the density gradient of the ambient medium seems to
be close to the projection plane, and hence will make the projection
effect weak according to the Petruk (2001) model. 

\subsubsection{Radiative rim}

%First, we note that the gas temperatures detected on the border
%are much higher than $10^{6}\K$, so the postshock gas at the rim
%has not suffered severely from radiative cooling.
%Therefore the whole remnant is still is the adiabatic stage.

Though there is no report in the available literature of the presence
of an HI shell around the remnant like that in W44, another thermal
composite SNR (Koo \& Heiles 1995), the diffuse filamentary near-infrared
[Fe II] and the mid-infrared 12-18 $\mu$m [Ne II] and [Ne III] emission
along the NW radio shell provide some evidence for a radiative cooling at
the rim (Reach, Rho, \& Jarrett 2001).
As viewed in Fig.1, the volume of the remnant as delineated by the
radio shell (including the regions of the near- and mid-infrared emission)
is basically filled with X-ray emitting gas.
%as a contrast to W44, where the X-ray
%emitting gas resides in only a small part interior to the radio shell
%(Giacani et al.\ 1997).
%In 3C~391, for instance, 
Several examined portions of X-ray emission
on the borders, such as the
NE rim (region \#10), the northern rim (region \#11), the western
rim (region \#12), the SW faint rim (region \#16), and the 
SE rim (region \#13), which should be dominated by the postshock
gas of the supernova blast wave (and are indeed located at the edge
of the radio shell), are found to be at temperatures
higher than $5\E{6}\K$.
This indicates that a considerable amount of gas at the blast shock has
not yet suffered heavy radiative cooling, and is capable of generating
considerable X-ray emission along the remnant's border.
% as observed in the \Chandra\ images (Fig.1).
If the gas density of these regions,
$\sim1$--$3f^{-1/2}\du^{-1/2}\cm^{-3}$, 
is approximated as the postshock density,
then the preshock intercloud medium (ICM) has a density about 
$0.2$--$0.7f^{-1/2}\du^{-1/2}\cm^{-3}$.

According to the infrared spectroscopic study, the ambient environment
in the NW part has complicated physical properties,
with at least three typical preshock components (Reach \& Rho 2000).
For the atomic component implied by the [OIII] and [NIII] emission,
the preshock density $\no<1\cm^{-3}$ and the shock velocity
$\vs\sim500\km\ps$;
for the moderate-density molecular component revealed by the [OI], [SiII],
and [FeII] emission, the gas density $n_{cl}\sim10^2\cm^{-3}$
and the shock velocity $v_{cl}\sim10^2\km\ps$;
and for the dense clumps giving rise to H$_2$, OH, CO, CS, and H$_2$O
molecular emission, the gas density $\sim10^{4}\cm^{-3}$ and the shock velocity $\sim20\km\ps$.
Thus there may be two possibilities about the coexistence of the radiative 
filaments and the X-ray emitting gas along the NW rim:
1) the cooling shell may be in the early stage of its formation; or
2) the radiative filaments may be caused by the propagation of the
shock waves into some regions of higher density % than the average density
in the molecular cloud so that these regions of postshock gas
have cooled down severely while most of the gas at the rim is still hot.

In the first case, we note that the bright [FeII] emission coincides
precisely with the brightest radio bar along the western border, which is
regarded as the interface between the remnant and the molecular cloud
(Reach, Rho, \& Jarrett 2001), and that, as we find here,
the soft X-ray brightened slab along the west border (in region \#12)
appears to be very close to, but slightly behind, the radio bar.
The shell formation time can be estimated according to
$t_{\rm shell}\approx5.3\E{4}(\Eu^{3/14}/\no^{4/7})\yr$ (Cox et al.\ 1999).
If the mean preshock hydrogen nucleus density is $\no\sim30\cm^{-3}$
(twice the molecular density
$\langle\nHH\rangle$ $\sim10$-$20\cm^{-3}$ given above),
$t_{\rm shell}\approx7.6\E{3}\Eu^{3/14}\yr$;
if the preshock density is $\no\sim100\cm^{-3}$ for the moderately
dense cloud, $t_{\rm shell}\approx3.8\E{3}\Eu^{3/14}\yr$.
The remnant's age would not be much larger than these estimates.

In the second case, the blast shock propagates in an inhomogeneous
medium.
After a blast shock impacts a cloud, a transmitted shock
(at velocity $v_{cl}$) is expected to move into the cloud.
The pressure balance between the shocked cloud and the shocked ICM
gives $n_{cl}v_{cl}^2=\beta n_{icm}\vs^2$, where factor
%$n_{cl}$ is the hydrogen nucleus density in the clouds,
$\beta\sim1$ for a shock
interaction with small clouds (McKee \& Cowie 1975) and 
$\beta\approx4.4$ and 6 for a strong shock hitting a rigid cloud plane
for $n_{cl}/\no=100$ and $\infty$, respectively (Zel'dovich \& Raizer 1967).
The moderate-density ($n_{cl}\sim100\cm^{-3}$) cloud may, after
it is shocked (with $v_{cl}\sim100\cm^{-3}$), be responsible
for the filamentary near- and mid-infrared emission
and the tenuous atomic component ($n_{icm}<1\cm^{-3}$) may,
after shocked, correspond to
the postshock X-ray emitting gas detected by \Chandra.
For the western X-ray slab ($kT\sim0.6\keV$) along the radiative filaments,
the velocity of the blast wave in the hot gas is
$\vs=(16kT/3\mu\mH)^{1/2}\sim700\km\ps$
(where the mean atomic weight $\mu=0.61$),
and the preshock density is $n_{icm}=\nH/4\sim0.5\cm^{-3}$.
Thus we have $\beta\sim4.1f^{1/2}\du^{1/2}$,
which suggests that the SNR shock wave hits the plane surface
of a dense cloud.
This is consistent with the flat morphology of the NW radio bar.

\subsubsection{Thermal conduction}
%As a comparison with SNR W44,

The spatially-resolved spectral analysis demonstrates
that the hot gas over most interior regions of the SNR is
at similar temperatures of around 0.5-$0.6\keV$.
Such a uniform distribution of the gas temperature inside the remnant
is not consistent with
the canonical Sedov distribution in which the inner temperature
is higher than that near the rim.
On the other hand, the X-ray morphology of this remnant is not
limb brightened, and this is inconsistent with the Sedov
solution as well.
Cox et al.\ (1999) explain this sort of morphology (as 
observed in W44) in terms of the effect
of thermal conduction of the interior gas, which could prevent
the formation of a hot vacuous cavity described by
the adiabatic Sedov solution and provide enough material in the
center to make it bright in X-rays.

The thermal conduction scenario could apply to 3C~391 to some extent.
If the suppression of conduction by magnetic fields 
can be ignored,
the conduction timescale would be
$t_{\rm cond}\sim \nel k\ell^2/\kappa$,
where $\ell$ denotes the linear scale of the temperature gradient
%is taken to be of order the remnant's mean radius $7\parsec$ and
and the conductivity is given by
$\kappa=1.84\E{-5}T^{5/2}/\ln\Lambda\ergs\ps\K^{-1}\cm^{-1}$
with the Coulomb logarithm
$\ln\Lambda=29.7+\ln\,\nel^{-1/2}(T/10^6\K)$ (Spitzer 1962).
If $\ell$ is taken to be of order the remnant's mean radius $7\parsec$,
the timescale would be $t_{\rm cond}\sim 4.7\E{4}\yr$,
much larger than the remnant's inferred age.
It would, however, be much smaller than this in the early stage of
the evolution in which $T$ would be higher and $\ell$ would be smaller.
If we look at small spatial scales of order the separation between clumps
(typified by $1'\sim2.3\parsec$),
the timescale would be $5.2\E{3}\yr$,
%just slightly larger than the remnant's age (see below),
comparable to the remnant's age (see \S3.1.4.),
and hence thermal conduction may play a role in smoothing of the
interior temperature profile.
We note that X-ray clumps are also present in W44
(Shelton, Kuntz, \& Petre 2004) where thermal
conduction may be an efficient mechanism.
The fact that the hot gas in 3C~391 is near ionization equilibrium
suggests that there may be relatively little
newly shocked material and is consistent with the idea of some
mechanism smoothing the properties.
More significantly, 
the presence of the radiative filaments along the NW border is
also in favor of the thermal conduction model.

There are, however, disagreements in the distribution of physical
properties between the observation and the model interpretation. 
In the thermal conduction scenario
in which a centrally brightened X-ray morphology is reproduced
(Cox et al.\ 1999, Shelton et al.\ 1999),
the central pressure is about 0.3 of that at the edge
and the central density would be about much lower
(0.13 times at radiative shell formation) than the preshock ICM density,
but this is not consistent with the case of 3C~391.
In fact, this remnant has a uniform temperature around
0.5-$0.6\keV$ throughout
and the inner mean density $\sim3f^{-1/2}\du^{-1/2}\cm^{-3}$
(regions \#14 and \#15) is not smaller than the hot gas density
$\sim1$--$3f^{-1/2}\du^{-1/2}\cm^{-3}$ along the periphery.
In the model the temperature would be expected
to decrease gradually away from the inner portions to the outer.
Unlike W44 in which the gas temperature is found to drop by a factor
$\lsim2$ between the center and $\sim6\parsec$ from the center
(Shelton et al.\ 2004),
in 3C~391 our measurement does not suggest the temperature decrease
even if the line-of-sight projection is taken into account.
The inner temperature is not higher than
the gas temperature in most portions along the rim, and highest
temperatures appear at a location on the border (region \#10, the NE rim).

\subsubsection{Cloudlet evaporation}

The phenomenon of low temperature in the inner portions
could result from substantial cloudlet evaporation
because part of the thermal energy is deposited
to the gas evaporated from the cold cloudlets.
The relatively uniform distribution of temperature (even with slightly
lower values at the center) is actually expected by the
White \& Long (1991) cloudlet evaporation model for model parameters
$\tau\rightarrow\infty$ and $C/\tau\gsim3$ (see Fig.2 and Fig.4 therein),
where $\tau$ is the ratio of the cloud evaporation timescale to the
SNR's age and $C$ is the ratio of the mass in the cloudlets
to the mass of ICM.
The ratio between the mean density ($\sim2f^{-1/2}\du^{-1/2}\cm^{-3}$)
and the density along the border ($\sim1$--$3f^{-1/2}\du^{-1/2}\cm^{-3}$)
 is basically
consistent to that predicted in the evaporation model.
In fact, the highly clumpy structure unveiled in this observation
lends support to the conjecture that the ambient molecular cloud is
inhomogeneous.
The cloudlets engulfed by the supernova blast wave can act as a large
reservoir of interior gas by gradual evaporation. This could also
explains why the mean density of the hot gas
inside the remnant (as mentioned above)
is much lower than that of the ambient cloud gas.
The association between the thermal composites and
the OH maser emission in molecular clouds
is strongly suggestive of the role
played by clouds in the SNRs (Yusef-Zadeh et al.\ 2003).

With a little adjustment of the model parameters used in the \ASCA\
study of 3C~391 (Chen \& Slane 2001, see Table 6 therein), we here adopt
$C/\tau\sim$3--6.
Using the temperature measured from the entire remnant (region \#17)
$kT_X\approx0.56\keV$ as the average temperature,
the White \& Long (1991) model would predict a
postshock temperature 0.46--$0.64\keV$,
consistent with most of the postshock temperatures measured from
the spectral analysis.
%Then the average temperature of the entire remnant $kT_X\approx0.56\keV$
%(measured using region \#17)
%would correspond to the postshock temperature 0.46--$0.64\keV$ according
%to the White & Long (1991) model,
%and is consistent with the postshock temperatures obtained from
%the spectral analysis.
The velocity of the blast wave is
$\vs=(16k\Ts/3\mu\mH)^{1/2}\sim620$--$730\km\ps$
(where the mean atomic weight $\mu=0.61$).
Adopting a mean radius of the remnant $\rs=3'\ru\sim7.0\ru\du$,
we estimate the dynamical age of the remnant:
$t=2r_{s}/5\vs\sim3.7$--$4.4\E{3}\ru\du\yr$.
Because our \Chandra\ observation can measure the postshock
temperature of 3C~391, this age estimate should be more
accurate than those obtained from \ROSAT\ and \ASCA\ observations
using the same model,
and the age of remnant is somewhat smaller than previously thought.
%%Though we have performed spectral analysis for the postshocks
%%regions and obtained the volume emission measure,
%The estimate of the gas density would be somewhat artificial
%since the real volume in each defined region occupied by
%the X-ray emitting gas is difficult to determine because of
%projection effects.
%As a raw estimate, we assume that the circular regions correspond
%to spheres, the elliptical regions to oblate spheroids, and the
%rectangular regions to shell structures, and thus the volume
%emission measures of the defined regions on the border give
%postshock densities of $\sim1$--$3f^{-1/2}\du^{-1/2}\cm^{-3}$
%except the SE and eastern regions \#1 and \#2 containing
%very bright knots (with density $\gsim10f^{-1/2}\du^{-1/2}\cm^{-3}$).
%The ratio between this estimate and the mean density
%($\sim2f^{-1/2}\du^{-1/2}\cm^{-3}$) given above is basically
%consistent to that predicted in the White \& Long model.
%So the density of the preshock ICM is
%$\no\sim0.2$--$0.7f^{-1/2}\du^{-1/2}\cm^{-3}$.
The preshock gas density ($\no\sim0.2$--$0.7f^{-1/2}\du^{-1/2}\cm^{-3}$)
estimated above could be regarded as the preshock ICM density.
An alternative estimate of the preshock ICM density is obtained
%from the remnant's X-ray luminosity ($L_X=2.3\E{36}\du^2\ergs\ps$):
from the remnant's X-ray luminosity ($L_X=3.5\E{36}\du^2\ergs\ps$):
%$\no\sim0.06$--$0.3\cm^{-3}$
$\no\sim0.07$--$0.4\cm^{-3}$
by use of eq.(21) in White \& Long's (1991) model.
According to this model, the explosion energy of the supernova
remnant is
$E=[16\pi(1.4\no\mH)/25(\gamma+1)K](\rs^5/t^2)
\sim0.3$--$1.4\E{51}\ru^3\du^3(\no/0.3\cm^{-3})\ergs$
%\sim2.2$--$9.0\E{50}\ru^3\du^3(\no/0.2\cm^{-3})\ergs$
(where the adiabatic index $\gamma=5/3$ and the ratio of
thermal to kinetic energy $K\sim0.385$--0.132).
%Alternatively, the thermal energy inside the remnant is
%$E_{\rm th}\sim2.3(3kT/2)V\sim2.9\E{50}\du^3\ergs$;
%in the adiabatic stage, $E\sim E_{\rm th}/0.71\sim4\E{50}\du^3\ergs$
%{(\bf McKee \& Ostriker 1977)}.

The application of the evaporating cloudlets model to 3C~391
may raise a few questions, which are discussed below and need further
clarification.

First, the remnant ages may be systematically underestimated
by the White \& Long model for a given explosion energy.
W44 is a typical example, where an age of $\sim7\E{3}\yr$ estimated from
the evaporation model (Rho et al.\ 1994) is smaller than the spindown
time scale $\sim2\E{4}\yr$ of the associated pulsar
(Wolszczan, Cordes, \& Dewey 1991).
The age derived above for 3C~391 ($\sim4\E{3}\ru\du\yr$)
using the evaporation model may possibly be an underestimate.
If a Sedov evolution is adopted, the postshock temperature
$k\Ts\sim0.6\keV$ would correspond to %shock velocity $\sim700\km\ps$
a similar age $2\rs/5\vs\sim4\E{3}\ru\du\yr$;
as a comparison, Reynolds \& Moffett (1993) present an estimate
$\sim1.7\E{4}\Eu^{-1}\yr$ for a cloud density $\no\sim100\cm^{-3}$.
Considering the possibility of the SNR being in the radiative phase,
Chen \& Slane (2001)
obtain an age $\sim1.9\E{4}\Eu^{31/42}(\ru\du)^{10/3}\yr$ for a preshock
density $\no\sim30\cm^{-3}$.

%While the cloud evaporation scenario is favored by
%the highly clumpy gas structure, the modest radial gradients
%of temperature and density, and the much lower hot gas density than the
%ambient cloud density,
%it meets a few questions that are discussed below and need further
%clarification.

Second, the cloudlet evaporation scenario needs a model parameter
$\tau\rightarrow\infty$ (with $C/\tau\gsim3$),
and thus the influence of cloud evaporation
on the properties of the interior gas seems possibly negligible.
Similar parameters are required to explain the X-ray morphology of
another thermal composite MSH11-61A observed with \ASCA\ (Slane et al.\ 2002).
In contrast, hydrodynamic simulations (e.g., Xu \& Stone 1995, using
a density contrast $\chi=10$)
show that clouds tend to get disrupted soon after passing through
a strong shock.
In our observation of 3C~391, however, the SE bright knots (in region \#1)
appear compact and survive the shock passage.
The ratio of the hot gas density of these knots to the mean density of
the entire remnant is about 10 (see Table 2),
so it would be reasonable to assume that 
the cloud-to-ICM density contrast $\chi$ is much higher than this number
or to assume a high mass contrast $C$.
The other X-ray knots interior to the remnant may also be the dense
regions that have not been destroyed by the blast shock.
The denser clouds could be expected to survive longer time, and still
a large amount of dense matter expected to have not been evaporated
from cloudlets yet.
Infrared observations of 3C~391 have detected dense ($\gsim10^4\cm^{-3}$)
molecular clumps in the southwest ``broad molecular line'' region,
each H$_2$ clump being of size $\sim0.1\parsec$
(Reach \& Rho 1999, Reach \& Rho 2000, Reach et al.\ 2002).
(As a comparison, the compact SE knots in region \#1
are of size $\sim0.5\parsec$.)
It is not impossible that the X-ray knots contain similar dense cloudlets.

Third, if the bright knots stands for the newly evaporated cloud
material, it should have been seen cooler than the mean temperature. 
Nonetheless, because the evaporated gas would be soon heated up to the
environmental temperature about 2 times the cloud radius from the cloud 
(Cowie \& McKee 1977), the spectral properties of the cooler component
could be smoothed out if the knots are much larger than the cold cloudlets
(for instance, 0.5 pc versus 0.1 pc as described above).

Fourth, the newly evaporated gas should have a low ionization age.
The knots in regions \#2, \#6, \#7, \#8, and \#9 marginally satisfy this
condition ($\gsim10^{11}\s$), especially if the evaporated gas had been
pre-ionized by the ultraviolet photons and/or
the transmitted shock wave, while those in region \#1 has a high
ionization age ($>10^{12}\s$). The latter case does not satisfy the
condition, unless, again, the low ionization age could be smeared
by the surrounding gas if the cold clouds are very small.

Concluding this section, we suggest that three mechanisms, radiative
rim, thermal conduction, and cloudlet evaporation, all play roles in
the inner brightened X-ray morphology of 3C~391.
By comparison, the cloudlet evaporation can largely explain
some important properties such as
the highly clumpy structure, the uniform temperature and density
distribution almost all over the remnant,
and the much lower X-ray emitting gas density than the mean ambient
cloud density.
Nonetheless, even cloudlet evaporation is driven by saturated conduction.
It is probably true in many thermal composite remnants, and in 3C 391 as well,
that the morphology arises from a
combination of expansion into an inhomogeneous medium and heat
conduction on various scales.

\subsection{Maser spots}
As in many other thermal composites, hydroxyl radical 1720 MHz maser
spots are found within 3C~391.
The two maser spots are located on the radio shell (Frail et al.\ 1996),
one of which is here found to be harbored in the NE rim
(region \#10) where X-ray emission is modestly enhanced at a high
plasma temperature ($\sim0.8\keV$) and
the other of which is in the southwest region (\#16) where the
X-ray emission is faint (see Fig.1).

The OH (1720 MHz) maser spots are believed to be 
due to C-type shock collisional excitation in the clumpy
molecular clouds, with cloud temperature and density conditions
$50\le T\le125\K$ and $10^4\le n_{{\ssst\rm H}_{2}}\le5\E{5}\cm^{-3}$
(Elitzur 1976; Lockett, Gauthier, \& Elitzur 1999).
The production of OH molecules is enhanced by the X-ray emission
(with an ionization rate $\gsim10^{-16}\ps$) behind the C-shock
(Wardle 1999).

According to Wardle's model formula, the X-ray induced ionization rate
in the postshock regions of 3C~391 is about $2\E{-15}\ps$
(using the X-ray luminosity $L_X\sim2.3\E{36}\ergs\ps$ and
the mean radius $\sim7\parsec$), well above the rate needed for
the OH enhancement.
Let us take the X-ray enhanced region \#10 and X-ray faint region \#16
as oblate spheroids.
Thus the volume in region \#10 is $\sim6.3\du^3\parsec^3$
and the mean gas density is $\sim2.3f^{-1/2}\du^{-1/2}\cm^{-3}$.
So, the pressure in the hot intercloud gas,
$p_{\rm icm}\approx2.3\nH kT\sim6.7\E{-9}f^{-1/2}\du^{-1/2}\ergs\cm^{-3}$.
Similarly, for region \#16, the volume is $\sim240\du^3\parsec^3$,
the mean gas density is $\sim1.1f^{-1/2}\du^{-1/2}\cm^{-3}$, and
the gas pressure is
$p_{\rm icm}\sim2.1\E{-9}f^{-1/2}\du^{-1/2}\ergs\cm^{-3}$.
Such gas pressures are consistent with the that estimated for the 
maser portions,
$p_{\rm cl}=n_{{\ssst\rm H}_{2}}kT\sim0.07$--$8.6\E{-9}\ergs\cm^{-3}$.
The far-infrared H$_{2}$O and OH emission lines are found
around the SW maser site. They are
consistent with the passage of shock wave through dense clumps
and the postshock gas is estimated to have a density $\sim2\E{5}\cm^{-3}$
and temperature 100--1000 K (Reach \& Rho 1998).
Such density and temperature are similar to the condition
for the maser production, with a pressure range
$p_{\rm cl}\sim2.8\E{-9}$--$2.8\E{-8}\ergs\cm^{-3}$,
comparable to the above pressure estimates for the SW maser region.

\subsection{Stellar remnant?}
The association of 3C~391 with a dense molecular cloud makes it
%possible that 3C~391 is a remnant of the explosion of a massive
%progenitor star.
possible that 3C~391 is the interstellar remnant of the supernova
explosion of a massive star, which may also leave behind a compact
star as a result of gravitational core collapse. It is 
thus important to see if there is any evidence
for this stellar remnant.
%The gravitational core collapse of the massive progenitor
%should have left behind a compact star,
%so it would be important to seek its potential stellar remnant.
We do locate numerous point-like sources within the boundary
of the remnant (for example, evident point sources
\#22 and \#26 right in the center, \#24 and \#32
in the south, etc.) (see Fig.1d and Table 1).
However, spectral analysis for these sources cannot be carried out
because of the small number of counts.

A bright unresolved source (source \#17) appears
near the NW rim and seems to be located at the north apex
of the NW diffuse hard emission (the blue patch in Fig.1a and Fig.1b).
The spectrum of this source extracted from region \#5
(with 158 net counts) is better described with a powerlaw than
with a bremsstrahlung or a blackbody (see Fig.4 and Table 3).
The hydrogen column density obtained is similar to those of
other portions of this remnant, so the possibility of the association
of this source with the remnant cannot be ruled out.
If this source is at the distance of 3C~391,
its unabsorbed X-ray luminosity is $\sim7.0\E{32}\du\ergs\ps$.
The steep photon index of the possible powerlaw model
2.5-4.5 is higher than the typical
indices 1.2-2.2 of active galactic nuclei (Turner \& Pounds 1989),
but is similar to those of the some ``compact central objects''
within SNRs (e.g.\ Pavlov et al.\ 2002).
If the association with the gaseous remnant is true, the stellar
remnant may have moved away from its explosion site.
If we follow Reynolds \& Moffett's
assumption that the explosion site is located at the center
of the NW half, then the source would have displaced 2' toward
the border. Taking an age of $\sim4\E{3}\yr$, then the displacement
velocity would be $\sim1100\km\ps$.
Such a transverse velocity is quite high, but not unreasonable;
pulsar proper motion studies have provided
evidence that a small population
of neutron stars have velocities in excess of $\gsim1000\km\ps$
(Lai 2003). For example, the pulsar B2224+65 in the Guitar
nebular pulsar has a velocity $\sim800$--$1600\km\ps$
(Chatterjee \& Cordes 2004).
Since the age used may be an underestimate, a higher remnant age
would allow for a lower velocity.

%If the unresolved source is an offset pulsar, it is interesting
%to search for the emission from a nearby pulsar wind nebula.
%A distinct extended region of hard emission is indeed seen near the NW
%source (the blue patch mentioned in \S2.1),
%appearing to be a possible trailing region of the
%source.
If the unresolved source is an offset pulsar, a trailing pulsar
wind nebula might be expected to be present to the south, roughly
in the hard emission region (the blue patch mentioned in \S2.1).
However, the spectrum of the diffuse gas of the blue patch (region \#4)
is well described by a thermal model (Table 2).
If a power law component is added to fit the spectrum, the photon
index would be $7.7^{+2.3}_{-2.2}$, too steep to be physically true.
Since the spectrum of region \#4 is very week below 1.3 keV (Fig.2)
and the hydrogen column density is very high ($\sim4.1\E{22}\cm^{-2}$),
this hard region seems to be the diffuse hot gas suffering
heavy absorption in soft X-rays.

\section{Conclusion}
We observed the thermal composite supernova remnant 3C~391 for
60 ks using \Chandra\ ACIS-S detector and carried out a detailed
spectro-spatial X-ray analysis which results in following
conclusions.
\begin{enumerate}
%The X-ray emission of this remnant is found to be highly clumpy.
\item The southeast-northwest elongated morphology is similar to that
previously found in radio and X-ray studies.
Faint shell-like X-ray structures appear along the northern and NE
boundary and a faint slab appears at the west rim,
while on the SW side the diffuse gas seems to expand out of
the radio boundary.
The X-ray emission is hardened in the NW, consistent with the
increase of hydrogen column density from SE to NW,
and all the images and spectra show compellingly that the intervening
column density is the primary influence on the broad band X-ray
appearance of the remnant.

\item This observation unveils a highly clumpy structure in the remnant.
The spatially resolved spectral analysis for the small-scale features
%yield normal abundance. except the clumps as protrusions on the SE border
%that may have slightly elevated metal abundances.
%The spectral analysis also
found that the gas
is generally of normal metal abundance and
has approached or basically reached ionization equilibrium.

\item Three mechanisms, radiative rim, thermal conduction,
and cloudlet evaporation, may all play roles in the X-ray visage
of 3C~391 as a ``thermal composite" SNR,
but there are difficulties with each of them in explaining
some physical properties.
By comparison, the cloudlet evaporation model can largely explain
some important properties such as
the highly clumpy structure, the uniform temperature and density
distribution almost all over the remnant,
and the much lower X-ray emitting gas density than the mean ambient
cloud density.
% provides the first
%a strong support for the
%cloudlet evaporation model to account for the X-ray emission of
%3C~391 as a ``thermal composite" SNR,
%while the temperature distribution is inconsistent with the
%interpretations invoking thermal conduction and/or a radiative
%shell, and 
The enhanced emission
in both the SE and NW halves as well as the hydrogen column density
gradient does not support the projection effect model.
The directly measured postshock temperature implies a young
age, $\sim4\E{3}\yr$, for the remnant.

\item The postshock gas pressure derived from the NE and SW rims,
which harbor maser spots,
is consistent with the estimate for the maser regions.

\item An unresolved source on the northwest border is best fitted 
in spectrum by a power-law, and its position and absorption column density
might suggest a possibility for an association with 3C~391.

\end{enumerate}

YC and YS acknowledge support from NSFC grants 10073003 \& 10221001
and grant NKBRSF-G19990754 of China Ministry of Science and Technology,
POS acknowledges support from NASA contract NAS8-39073
and grant GO2-3081X,
and QDW acknowledges NASA LTSA grant NAG5-8935.
The authors thank an anonymous referee for helpful comments and advices.

\clearpage

\begin{deluxetable}{lrrrrr}
  \tabletypesize{\footnotesize}
  \tablecaption{{\sl Chandra} List of Point-like Sources on S3 Chip\label{acis_source_list}}
  \tablewidth{0pt}
  \tablehead{
  \colhead{Source} &
  \colhead{CXO} &
  \colhead{$\delta_x$} &
  \colhead{log(P)} &
  \colhead{Count Rate} &
  \colhead{HR} \\ 
  \colhead{No.} &
  \colhead{No.} &
  \colhead{($''$)} & &
  $(10^{-3}~{\rm cts~s}^{-1})$ \\
  \noalign{\smallskip}
  \colhead{(1)} &
  \colhead{(2)} &
  \colhead{(3)} &
  \colhead{(4)} &
  \colhead{(5)} &
  \colhead{(6)} 
  }
  \startdata
   1 & J184915.0-005651 &  0.3 &$ -15.0$&$     0.94  \pm   0.14$&               $ 0.86\pm0.09$        \\
   2 & J184916.1-005827 &  0.3 &$ -15.0$&$     0.43  \pm   0.09$&                          --        \\
   3 & J184916.2-005624 &  0.4 &$ -15.0$&$     0.47  \pm   0.10$&               $ 0.99\pm0.11$        \\
   4 & J184916.8-005841 &  0.5 &$ -11.2$&$     0.20  \pm   0.07$&                                       --       \\
   5 & J184917.0-005701 &  0.7 &$  -9.1$&$     0.24  \pm   0.07$&                                       --       \\
   6 & J184917.4-005710 &  0.4 &$ -15.0$&$     0.41  \pm   0.09$&                                       --    \\
   7 & J184917.4-005625 &  0.6 &$ -13.4$&$     0.30  \pm   0.08$&                                       --       \\
   8$^{\rm a}$ & J184917.5-005900 &  0.4 &$  -7.3$&$     0.19  \pm   0.07$&                                       --       \\
   9 & J184918.9-005252 &  1.0 &$  -9.1$&$     0.33  \pm   0.09$&                                       --          \\
  10 & J184919.0-005749 &  0.3 &$ -15.0$&$     0.33  \pm   0.08$&                                       --       \\
  11$^{\rm a}$ & J184920.4-005928 &  0.5 &$  -7.5$&$     0.13  \pm   0.06$&                                       --          \\
  12$^{\rm a}$ & J184921.1-005357 &  0.7 &$  -7.0$&$     0.53  \pm   0.13$&                            --          \\
  13$^{\rm a}$ & J184921.5-005413 &  0.6 &$  -7.0$&$     0.44  \pm   0.12$&                                       --          \\
  14$^{\rm a}$ & J184921.6-005355 &  0.6 &$  -9.3$&$     0.59  \pm   0.14$&                            --          \\
  15$^{\rm a}$ & J184921.9-005351 &  0.6 &$  -7.2$&$     0.48  \pm   0.13$&                                       --          \\
  16$^{\rm a}$ & J184922.0-005946 &  0.8 &$  -7.8$&$     0.14  \pm   0.06$&                                       --          \\
  17$^{\rm a}$ & J184922.3-005334 &  0.3 &$ -15.0$&$     2.11  \pm   0.21$&  $ 0.49\pm0.09$     \\
  18 & J184922.6-005458 &  0.5 &$  -9.9$&$     0.49  \pm   0.12$&               $ 0.92\pm0.13$        \\
  19$^{\rm a}$ & J184923.1-005736 &  0.5 &$  -8.7$&$     0.22  \pm   0.07$&                                       --       \\
  20 & J184923.7-005705 &  0.3 &$ -10.5$&$     0.27  \pm   0.08$&                                       --       \\
  21 & J184924.5-005447 &  0.5 &$  -9.3$&$     0.47  \pm   0.11$&                            --       \\
  22 & J184925.0-005628 &  0.2 &$ -15.0$&$     1.22  \pm   0.16$&               $ 0.57\pm0.11$     \\
  23 & J184925.5-005344 &  0.5 &$ -15.0$&$     0.64  \pm   0.13$&               --       \\
  24 & J184925.7-005749 &  0.2 &$ -15.0$&$     0.65  \pm   0.11$&               $ 0.87\pm0.11$        \\
  25 & J184926.1-010024 &  0.4 &$ -15.0$&$     0.46  \pm   0.10$&                          --       \\
  26 & J184927.0-005640 &  0.2 &$ -15.0$&$     1.16  \pm   0.15$&               $ 0.69\pm0.10$     \\
  27$^{\rm a}$ & J184927.0-005410 &  0.5 &$  -8.2$&$     0.66  \pm   0.16$&                            --       \\
  28$^{\rm a}$ & J184927.4-005349 &  0.5 &$  -9.5$&$     0.53  \pm   0.13$&                                       --          \\
  29 & J184927.8-005318 &  0.8 &$  -9.2$&$     0.46  \pm   0.11$&                            --       \\
  30$^{\rm a}$ & J184927.9-005330 &  0.7 &$  -8.2$&$     0.42  \pm   0.11$&                                       --          \\
  31$^{\rm a}$ & J184928.3-005734 &  0.4 &$  -8.9$&$     0.27  \pm   0.08$&                                       --          \\
  32 & J184929.1-005741 &  0.2 &$ -15.0$&$     0.97  \pm   0.14$&               $ 0.75\pm0.11$        \\
  33 & J184929.4-005951 &  0.6 &$ -15.0$&$     0.22  \pm   0.07$&                                       --       \\
  34 & J184929.5-005905 &  0.3 &$ -15.0$&$     0.52  \pm   0.10$&               $ 1.00\pm0.10$        \\
  35$^{\rm a}$ & J184931.3-005924 &  0.6 &$  -7.1$&$     0.22  \pm   0.08$&                                       --          \\
  36$^{\rm a}$ & J184932.7-005805 &  0.5 &$  -7.7$&$     0.32  \pm   0.09$&                                       --          \\
  37 & J184933.9-005406 &  0.9 &$ -10.0$&$     0.23  \pm   0.08$&                                       --          \\
  38 & J184934.4-005238 &  0.9 &$ -11.4$&$     0.56  \pm   0.12$&                            --       \\
  39 & J184937.0-005825 &  0.4 &$ -12.7$&$     0.58  \pm   0.12$&                            --          \\
  40$^{\rm a}$ & J184937.3-005438 &  1.3 &$  -8.3$&$     0.32  \pm   0.09$&                                       --          \\
  41$^{\rm a}$ & J184938.5-005848 &  0.4 &$ -13.1$&$     0.87  \pm   0.16$&               $ 0.78\pm0.14$     \\
  42$^{\rm a}$ & J184938.6-005629 &  0.5 &$ -10.2$&$     0.79  \pm   0.16$&               $ 0.86\pm0.15$        \\
  43$^{\rm a}$ & J184938.6-005845 &  0.4 &$ -10.1$&$     1.03  \pm   0.18$&               $ 0.78\pm0.13$     \\
  44$^{\rm a}$ & J184939.6-005820 &  0.4 &$ -12.2$&$     0.91  \pm   0.16$&               $ 0.64\pm0.14$     \\
  45 & J184939.7-005721 &  0.6 &$ -12.4$&$     0.42  \pm   0.10$&               $ 1.00\pm0.16$        \\
  46$^{\rm a}$ & J184939.7-005817 &  0.5 &$ -10.2$&$     0.74  \pm   0.15$&               $ 0.75\pm0.15$        \\
  47$^{\rm a}$ & J184939.8-005849 &  0.5 &$  -9.6$&$     1.39  \pm   0.20$&  $ 0.35\pm0.13$        \\
  48 & J184939.9-005532 &  0.4 &$ -15.0$&$     0.72  \pm   0.12$&                          --       \\
  49$^{\rm a}$ & J184940.0-005848 &  0.3 &$ -15.0$&$     1.77  \pm   0.21$&  $ 0.51\pm0.10$  \\
  50$^{\rm a}$ & J184940.0-005643 &  0.3 &$ -15.0$&$     1.93  \pm   0.23$&  $ 0.30\pm0.11$     \\
  51 & J184941.4-005441 &  0.9 &$ -10.9$&$     0.51  \pm   0.11$&               $ 0.86\pm0.14$        \\
  52 & J184944.7-005428 &  0.4 &$ -15.0$&$     2.95  \pm   0.25$&                          --       \\
\enddata
\tablecomments{ Column
(1): Generic source number.
(2): {\sl Chandra} X-ray Observatory (unregistered) source name,
  following the {\sl Chandra} naming convention and the IAU
  Recommendation for Nomenclature
  (e.g., http://cdsweb.u-strasbg.fr/iau-spec.html).
(3): Position uncertainty (1$\sigma$) in units of arcsec.
(4): The false detection probability P that the detected number
  of counts may result from the Poisson fluctuation of the local
  background within the detection aperture [log(P) smaller than
  -20.0 is set to -20.0].
(5): On-axis (exposure-corrected) source count rate in the
  0.3-7 keV band.
(6): The hardness ratio defined as
  ${\rm HR}=({\rm H-S})/({\rm H+S})$,
  where S and H are the net source count rates in the
  0.3--1.5 and 1.5--7~keV % for ACIS-S
  bands, respectively.
  The hardness ratios are calculated only for sources with
  individual signal-to-noise ratios greater than 4
  in the broad band (B=S+H), and only the values with
  uncertainties less than 0.2 are included.
}
  \tablenotetext{a}{\phantom{0} These sources are not removed
  for spectral analysis, either because of high $\log(P)$ (we define a
  threshold $>-9.0$) or high possibility of being clumps of
  diffuse emission (judged by eyes).}
  \end{deluxetable}
\clearpage

%\centerline{\begin{tabular}{cc|cccccc}
\begin{center}
\begin{deluxetable}{cc|cccccc|c}
\tabletypesize{\footnotesize}
\rotate
\tablecaption{VNEI fitting results with the 90\% confidence ranges
  and estimates of the gas density}
\tablewidth{0pt}
\tablehead{
\colhead{regions} & \colhead{net count rate} \vline &
\colhead{$\chi^{2}/{\rm d.o.f.}$} & \colhead{$\NH$} &
\colhead{$kT_{x}$} & \colhead{$n_e t$} &
\colhead{$f\nel\nH V/\du^{2}$\hspace{2mm}$^{\rm a}$} &
\colhead{$F^{(0)}(0.5$-$10\keV)$} \vline &
\colhead{$\nH/f^{-1/2}\du^{-1/2}$} \\
\noalign{\smallskip}
\colhead{} & \colhead{($10^{-2}$ cts$\ps$)} \vline & \colhead{} &
\colhead{($10^{22}\cm^{-2}$)} &
\colhead{(keV)} & \colhead{($10^{11}\cm^{-3}\,{\rm s}$)} &
\colhead{($10^{57}\cm^{-3}$)} & \colhead{($10^{-11}\ergs\cm^{-2}\ps$)}
\vline & \colhead{(cm$^{-3}$)}
}
\startdata
1  & $3.88\pm0.08$ & 90.5/65 & $2.8\pm0.1$ &
  $0.67^{+0.02}_{-0.04}$ & $>200$ & $2.55^{+0.27}_{-0.53}$ & 0.84 & 18\\
%1$^{\rm b}$ & $3.88\pm0.08$ & 80.9/61 & $3.1\pm0.4$ &
%  $0.67\pm0.05$ & $>200$ & $1.92\pm0.72$ & 1.50 \\
%      &    & \multicolumn{6}{c}{(
%  [Mg/H]=$1.94^{+3.56}_{-0.92}$, [Si/H]=$1.80^{+2.01}_{-0.59}$,
%   [S/H]=$1.68^{+1.47}_{-0.54}$, [Fe/H]=$3.00^{+8.14}_{-1.78}$)}\\
2  & $4.80\pm0.09$ & 87.0/79 & $2.7\pm0.1$ &
  $0.58^{+0.06}_{-0.05}$ & $5.1^{+4.6}_{-1.3}$ &
  $3.84^{+1.33}_{-0.89}$ & 1.31 & 11\\
3  & $2.49\pm0.07$ & 68.8/48 & $3.4\pm0.2$ &
  $0.56^{+0.06}_{-0.09}$ & $>6.4$ & $3.46^{+2.08}_{-0.90}$ & 1.05 & 6.4\\
%4$^{\rm c}$  & $2.10\pm0.06$ & 53.2/43 & $3.5^{+0.3}_{-0.2}$ &
%  $0.82^{+0.08}_{-0.05}$ & $>4.7$ & $1.50^{+0.65}_{-0.32}$ & 0.47 \\
4$^{\rm b}$  & $4.57\pm0.09$ & 108.9/79 & $4.1^{+0.3}_{-0.2}$ &
  $0.62^{+0.04}_{-0.05}$ & $>30$ & $9.42^{+3.10}_{-2.33}$ & 2.95 & 4.8\\
      &    & \multicolumn{6}{c}{(
  [Mg/H]=$1.12^{+0.13}_{-0.09}$, [Si/H]=$0.87^{+0.08}_{-0.05}$,
   [S/H]=$0.80^{+0.13}_{-0.09}$)} \vline \\
6  & $3.30\pm0.08$ & 65.2/58 & $2.7^{+0.2}_{-0.1}$ &
  $0.56^{+0.08}_{-0.06}$ & $5.0^{+3.8}_{-2.1}$ &
  $3.20^{+1.41}_{-0.98}$ & 0.97 & 6.1\\
7  & $3.31\pm0.08$ & 57.8/58 & $3.0^{+0.1}_{-0.2}$ &
  $0.54^{+0.05}_{-0.06}$ & $>5.5$ & $3.97^{+1.65}_{-0.98}$ & 1.20 & 6.8\\
8  & $3.08\pm0.07$ & 72.4/52 & $2.9^{+0.2}_{-0.1}$ &
  $0.63^{+0.08}_{-0.07}$ & $>3.4$ & $2.52^{+0.60}_{-0.65}$ & 0.95 & 5.4\\
9  & $2.23\pm0.06$ & 58.2/43 & $2.9\pm0.2$ & $0.63^{+0.07}_{-0.11}$ &
  $>2.7$ & $1.72^{+1.35}_{-0.41}$ & 0.62 & 4.5\\
10  & $2.14\pm0.06$ & 44.2/44 & $3.0^{+0.2}_{-0.1}$ &
  $0.79^{+0.14}_{-0.10}$ & $3.1^{+3.6}_{-1.4}$ &
  $1.16^{+0.52}_{-0.34}$ & 0.40 & 2.3\\
11  & $3.64\pm0.09$ & 96.9/72 & $2.8\pm0.1$ &
  $0.59^{+0.06}_{-0.04}$ & $>3.8$ & $3.17^{+0.38}_{-0.77}$ & 0.94 & 1.9\\
12  & $1.41\pm0.05$ & 49.8/32 & $3.7^{+0.6}_{-0.4}$ &
  $0.58^{+0.11}_{-0.12}$ & $>265$ & $2.28^{+3.19}_{-0.96}$ & 0.68 & 2.0\\
13  & $4.71\pm0.10$ & 107.0/82 & $3.0^{+0.2}_{-0.1}$ &
  $0.46^{+0.04}_{-0.03}$ & $>106$ & $8.23^{+2.73}_{-2.12}$ & 2.27 & 2.6\\
%14$^{\rm d}$  & $31.3\pm0.2$ & 265.1/179 & $2.7\pm0.1$ &
%  $0.56^{+0.02}_{-0.01}$ & $>7.7$ & $32.92^{+4.06}_{-3.84}$ & 7.13\\
%      &    & \multicolumn{6}{c}{(
%  [Mg/H]=$0.85^{+0.15}_{-0.12}$, [Si/H]=$0.79^{+0.08}_{-0.07}$,
%   [S/H]=$0.78^{+0.11}_{-0.10}$, [Fe/H]=$0.53^{+0.25}_{-0.18}$)}\\
14$^{\rm b}$  & $31.3\pm0.2$ & 273.6/180 & $2.9\pm0.1$ &
  $0.55\pm0.02$ & $>9.0$ & $35.00^{+3.75}_{-5.38}$ & 10.8 & 2.9\\
      &    & \multicolumn{6}{c}{(
  [Mg/H]=$1.12^{+0.13}_{-0.09}$, [Si/H]=$0.87^{+0.08}_{-0.05}$,
   [S/H]=$0.80^{+0.13}_{-0.09}$)} \vline \\
%15$^{\rm d}$  & $16.3\pm0.2$ & 234.3/162 & $3.3^{+0.2}_{-0.3}$ &
%  $0.55\pm0.02$ & $>359$ & $31.64^{+5.38}_{-4.82}$ & 6.17 \\
%      &    & \multicolumn{6}{c}{(
%  [Mg/H]=$0.64^{+0.27}_{-0.17}$, [Si/H]=$0.47^{+0.07}_{-0.06}$,
%   [S/H]=$0.63^{+0.11}_{-0.10}$, [Fe/H]=$0.46^{+0.43}_{-0.27}$)}\\
15$^{\rm b}$  & $16.3\pm0.2$ & 237.8/163 & $3.5\pm0.1$ &
  $0.54^{+0.02}_{-0.01}$ & $>23$ & $34.08^{+5.42}_{-4.20}$ & 10.2  & 2.8\\
      &    & \multicolumn{6}{c}{(
  [Mg/H]=$0.92^{+0.14}_{-0.13}$, [Si/H]=$0.53\pm0.05$,
   [S/H]=$0.64^{+0.12}_{-0.11}$)} \vline \\
%16$^{\rm d}$  & $6.08\pm0.13$ & $155.2/126$ & $3.0\pm0.4$ &
%  $0.53^{+0.05}_{-0.06}$ & $>239$ & $9.54^{+4.99}_{-2.60}$ & 1.95 \\
%      &    & \multicolumn{6}{c}{(
%  [Mg/H]=$0.70^{+0.54}_{-0.28}$, [Si/H]=$0.49^{+0.15}_{-0.11}$,
%   [S/H]=$0.54^{+0.23}_{-0.21}$, [Fe/H]=$0.52^{+0.96}_{-0.43}$)}\\
16$^{\rm b}$  & $6.08\pm0.13$ & $156.3/127$ & $3.2\pm0.2$ &
  $0.53^{+0.04}_{-0.06}$ & $>239$ & $10.0^{+5.42}_{-2.60}$ & 2.94 & 1.1\\
      &    & \multicolumn{6}{c}{(
  [Mg/H]=$0.93^{+0.28}_{-0.26}$, [Si/H]=$0.54^{+0.12}_{-0.11}$,
   [S/H]=$0.54^{+0.24}_{-0.21}$)} \vline \\
%17$^{\rm d}$  & $112.0\pm0.5$ & $693.1/295$ & $2.9\pm0.1$ &
%  $0.57\pm0.01$ & $12.1^{+10.7}_{-3.2}$ & $141^{+11}_{-10}$ & 29.9 \\
%      &    & \multicolumn{6}{c}{(
%  [Mg/H]=$0.70^{+0.11}_{-0.08}$, [Si/H]=$0.63^{+0.03}_{-0.04}$,
%   [S/H]=$0.68\pm0.05$, [Fe/H]=$0.52^{+0.15}_{-0.13}$)}
17$^{\rm b}$  & $112.0\pm0.5$ & $710.5/296$ & $3.1\pm0.1$ &
  $0.56\pm0.01$ & $>12.8$ & $150\pm10$ & 45.5 & 1.9\\
      &    & \multicolumn{6}{c}{(
  [Mg/H]=$0.97^{+0.07}_{-0.05}$, [Si/H]=$0.70\pm0.03$,
   [S/H]=$0.71^{+0.06}_{-0.05}$)} \vline & \mbox{}
\tablecomments{
%a: $f$ denotes the filling factor of the hot gas.\\
%b: In this case, abundances of Mg, Si, S, and Fe are thawed as free parameters.\\
%c: An unresolved source circled as ``region 5'' has been removed in the spectrum extraction.\\
%d: Thawing metal abundances apparently improves fitting.\\
}
\enddata
  \tablenotetext{a}{\phantom{0} $f$ denotes the filling factor of the hot gas.}
%  \tablenotetext{b}{\phantom{0} In this case, abundances of
%   Mg, Si, S, and Fe are thawed as free parameters.}
%  \tablenotetext{c}{\phantom{0} An unresolved source circled
%   as ``region 5'' has been removed in the spectrum extraction.}
  \tablenotetext{b}{\phantom{0} Thawing metal abundances apparently
   improves fitting.}
\end{deluxetable}
\end{center}
\clearpage

\begin{center}
\begin{deluxetable}{cc|cccccc}
\tabletypesize{\footnotesize}
%\rotate
\tablecaption{Spectral fitting results for the NW unresolved source with the 90\% confidence ranges}
\tablewidth{0pt}
\tablehead{
\colhead{regions} & \colhead{net count rate} \vline & \colhead{model} &
\colhead{$\chi^{2}/{\rm d.o.f.}$} & \colhead{$\NH$} &
\colhead{photon index} & \colhead{$kT$} & \colhead{$F^{(0)}(0.5$-$10\keV)$} \\
\noalign{\smallskip}
\colhead{} & \colhead{($10^{-3}$ cts$\ps$)} \vline & \colhead{} & \colhead{} &
\colhead{($10^{22}\cm^{-2}$)} &
\colhead{} & \colhead{(keV)} &
\colhead{($10^{-14}\ergs\cm^{-2}\ps$)}
}
\startdata
5 & $2.60\pm0.20$ & power & 4.08/4 & $2.6^{+1.4}_{-1.0}$ &
  $3.2^{+1.3}_{-0.7}$ & \nodata & 9.1 \\
\colhead{} & \colhead{(1--10 keV)}
\vline & bremsstrahlung & 4.86/4 & $1.8^{+0.9}_{-0.8}$ &
  \nodata & $2.2^{+1.4}_{-0.9}$ & 5.5 \\
\colhead{} & \colhead{}
\vline & blackbody & 6.95/4 & $0.6^{+0.9}_{-0.6}$ &
  \nodata & $0.75^{+0.15}_{-0.17}$ & 3.3 \\
%\tablecomments{
%a: $f$ denotes the filling factor of the hot gas.\\
\enddata
%  \tablenotetext{a}{\phantom{0} $f$ denotes the filling factor of the hot gas.}
\end{deluxetable}
\end{center}
\clearpage

\begin{center}
\section{Figure captions}
\end{center}

\figcaption[f1.eps]{
%X-ray images of 3C~391.
{\sl Panal a}: Tri-color X-ray image of 3C~391. The X-ray intensity
%in the 0.3-2.1, 2.1-2.6, and 2.6-10 keV bands are color coded in
in the 0.3-1.5, 1.5-3.0, and 3.0-7.0 keV bands are color coded in
red, green, and blue, respectively.
The overlaid 1.5 GHz radio contours are at 1.5, 3, 6, 9, 12, 18, 30,
and 45$\E{-3} {\rm~Jy~beam^{-1}}$ (from Moffett \& Reynolds [1994]).
{\sl Panel b}: Tri-color X-ray + radio image of 3C391.
%The 1.5 GHz emission is plotted in red, the 0.3-2.6 keV emission
%is in green, and 2.6-10 keV is in blue.
The 1.5 GHz emission is plotted in red, the 0.3-3.0 keV emission
is in green, and 3.0-7.0 keV is in blue.
%{\sl Panel c}: The broad band (0.3-10 keV) X-ray image
{\sl Panel c}: The broad band (0.3-7.0 keV) X-ray image
overlaid with 1.5 GHz radio contours (with the same levels described
for the top panel).
All the X-ray maps used here are exposure-corrected and
are adaptively smoothed to achieve a S/N ratio of 3
(using the CIAO program CSMOOTH).
The two cross labels in each panel denote the OH maser points
(Frail et al.\ 1996).
{\sl Panel d}: Diffuse emission from SNR~3C~391 in the broad band
0.3-7.0 keV. The image is smoothed adaptively with a Gaussian kernel,
which is adjusted to a count-to-noise ratio of 6.
%The contours are at
%3, 6, 13, 24, 50, 120, 220, and 400
%$\sigma$ above the background ($1.0 \times 10^{-2}
%{\rm~ct~s^{-1}~arcmin^{-2}}$;
%1$\sigma = 1.67 \times 10^{-3} {\rm~ct~s^{-1}~arcmin^{-2}}$).
The color is scaled in the range from (1.19--331.85)$\E{-2}
{\rm~ct~s^{-1}~arcmin^{-2}}$, logarithmically.
The location of the sources removed from the data before
the smoothing are marked by crosses.
All the regions used for extracting spectra are indicated in blue,
with cyan numerical labels.
The dashed lines denote the border of the S3 chip.
}
%\figcaption[f2.ps]{
%Narrow band 1.2-1.5, 1.7-2.0, and 2.3-2.6$\keV$
%(including Mg He$\alpha$, Si He$\alpha$, and S He$\alpha$, respectively)
%diffuse emission images ({\sl panel a, b,} and {\sl c} in sequence)
%overlaid with the dashed contours
%of 1.5 GHz radio emission (from Moffett \& Reynolds [1994]).
%The CIAO tool {\em csmooth} (with signal-to-noise
%ratio of 3) and exposure-correction have been applied, and
%the point sources have been removed.
%%The seven levels of solid contours are linear up to the maximum.
%The ten levels of solid contours are logarithmic between the maximum
%and the 4\% maximum brightness.
%The two plus signs in each panel denote the OH maser points
%(Frail et al.\ 1996).
%}
\figcaption[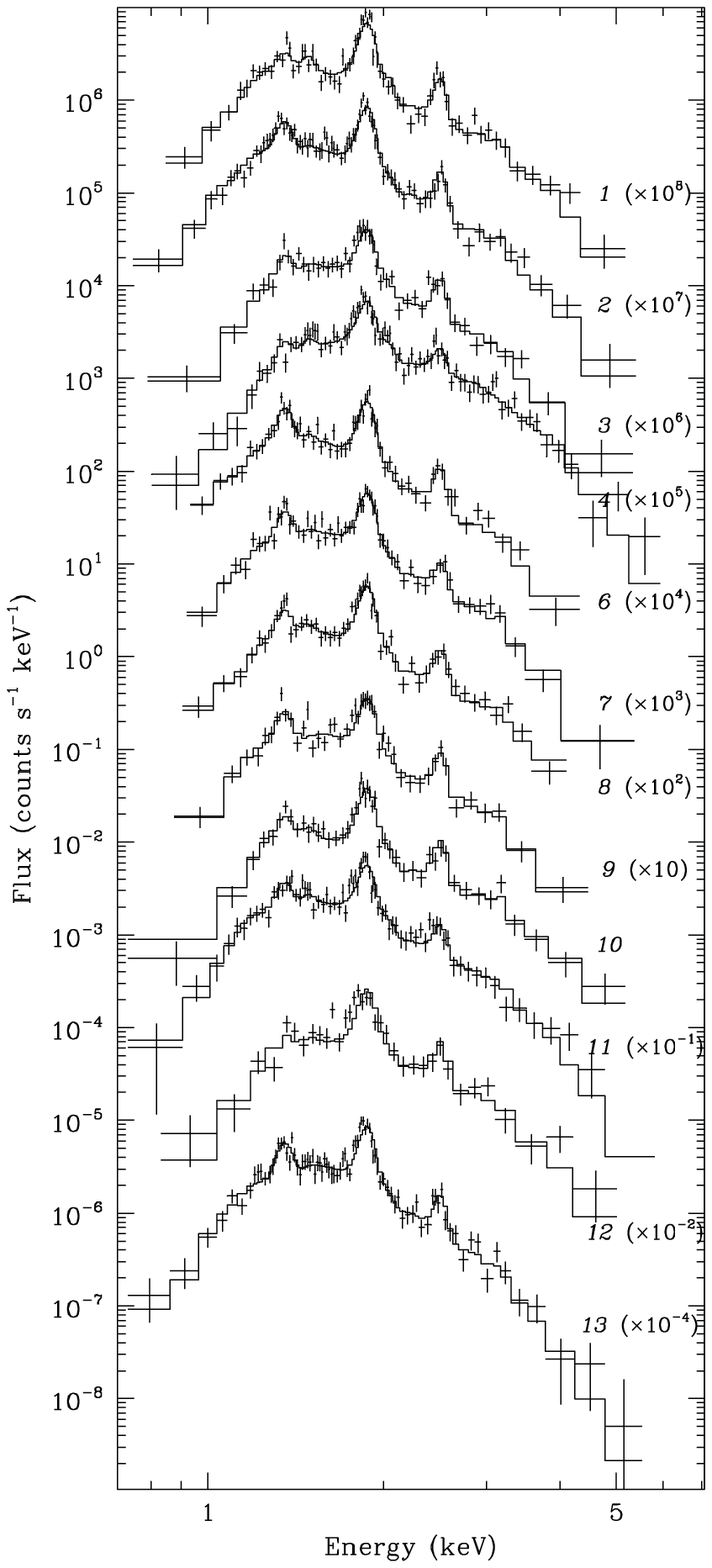]{
\Chandra\ ACIS spectra of 12 small scale emission regions
in 3C~391 fitted with the VNEI model.
The region numbers, together with the factors multiplying the
real fluxes, are labeled on the right side of the spectra.
}
\figcaption[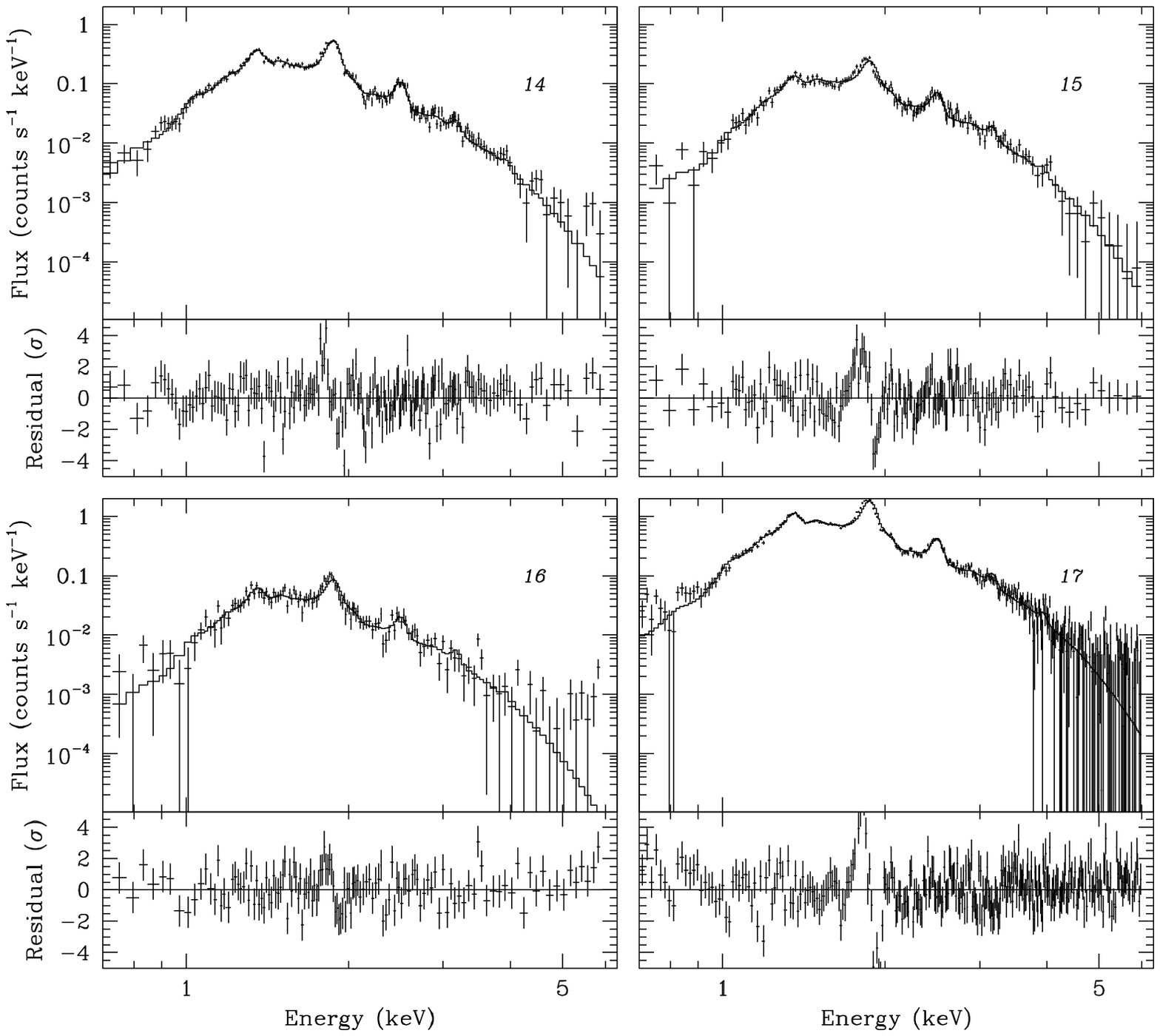]{
\Chandra\ ACIS spectra of the medium-scale emission regions
and entire remnant of 3C~391.
The region numbers are labelled in each panel.
}
\figcaption[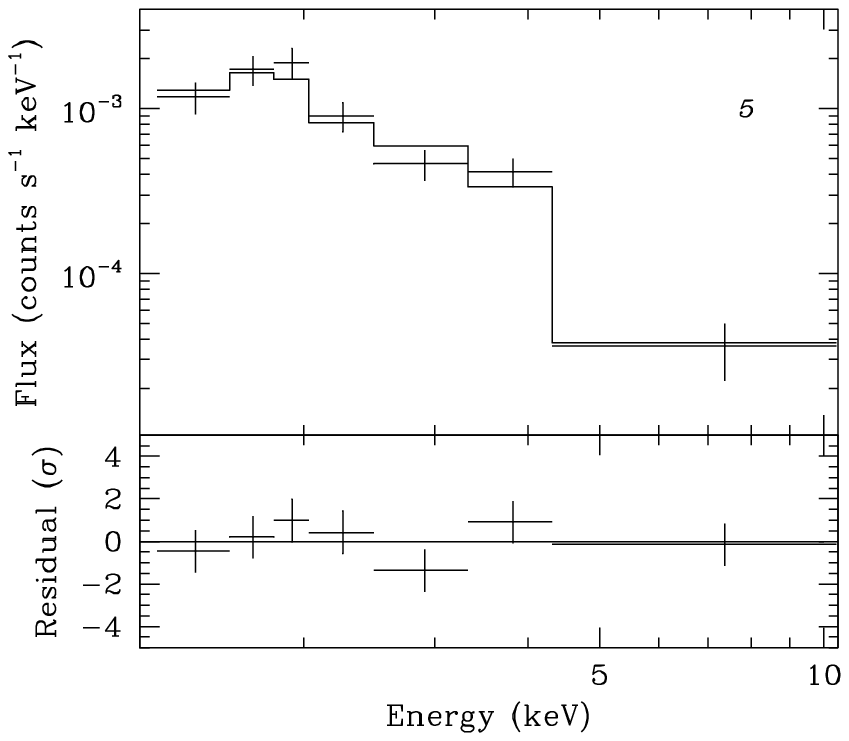]{
\Chandra\ ACIS spectrum of the NW unresolved source (region \#5)
fitted with the powerlaw model.
}
\clearpage

%\plotone{f1.eps}
Fig.1 (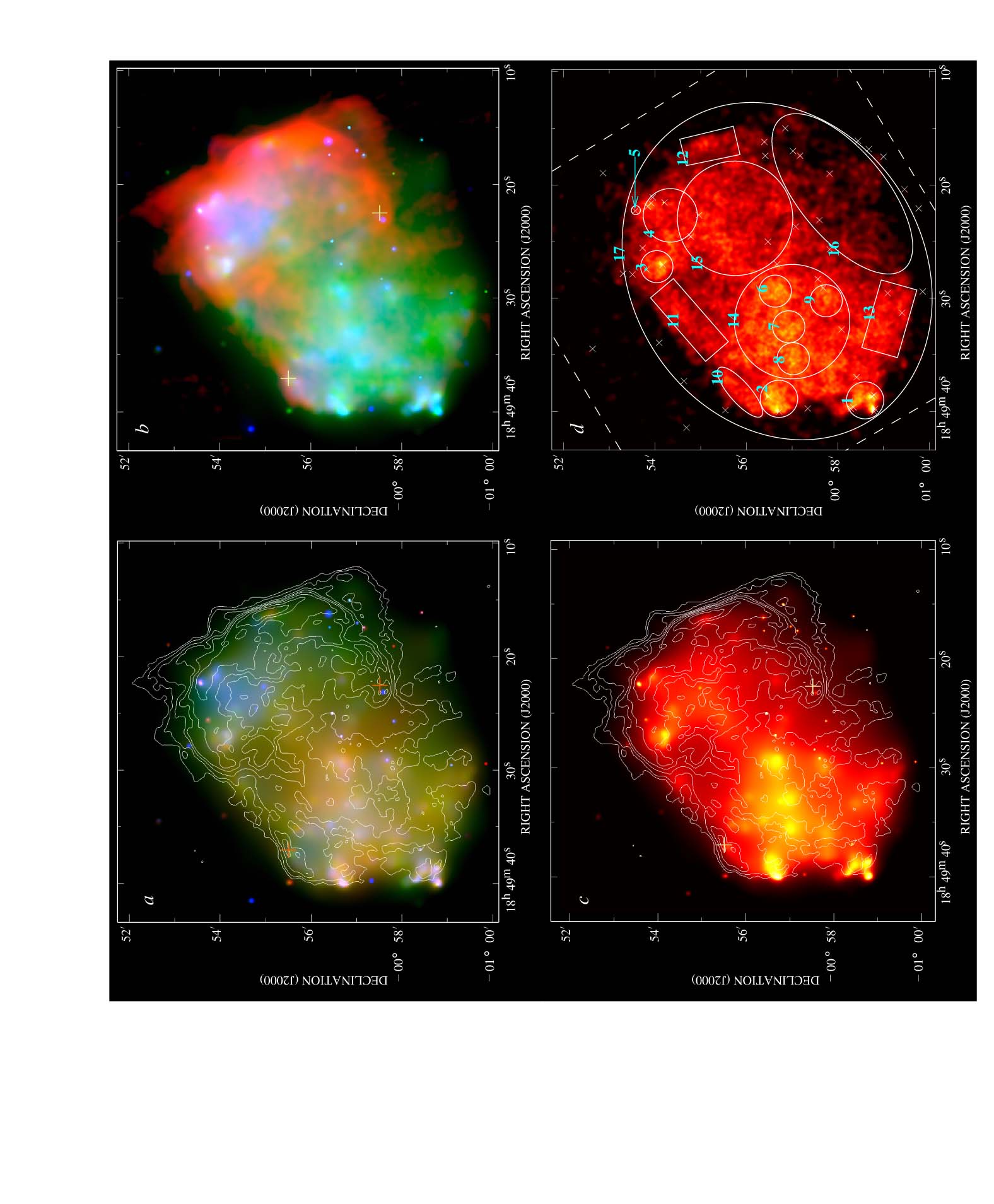)
\clearpage

\plotone{f2.eps}
Fig.2

\plotone{f3.eps}
Fig.3

\plotone{f4.eps}
Fig.4

%\plotone{f5.eps}
%Fig.5


\clearpage
\begin{thebibliography}{}
%\bibitem{} Aschenbach, B.\ 2002, in Proceedings of the 270 WE-Heraeus
%Seminar on ``Neutron Satrs, Pulsars and Supernova Remnants",
%eds.\ W.Becker, H.Lesch, \& J.Tr\"{u}mper, MPE Report 278, pp.\ 13-25
%\bibitem{} Aschenbach, B., Egger, R., \& Tr\"umper, J.\ 1995, Nature,
%373, 587
\bibitem{} Chatterjee, S. \& Cordes, J.M.\ 2004, \apjl, 600, L51
\bibitem{} Chen, Y. \& Slane, P.O.\ 2001, \apj, 563, 202
\bibitem{} Cowie, L.L.\ \& McKee, C.F.\ 1977, \apj, 211, 135
\bibitem{} Cox, D. P., Shelton, R. L., Maciejewski, W., Smith, R. K.,
Plewa, T., Pawl, A., \& R\'{o}zyczka, M. 1999, \apj, 524, 179
\bibitem{} Elitzur, M.\ 1976, \apj, 203, 124
\bibitem{} Frail, D.A., Goss, W.M., Reynoso, E.M., Giacani, E.B.,
Green, A.J., \& Otrupcek, R. 1996, \aj, 111, 1651
%\bibitem{} Giacani, E. B., Dubner, G.M., Kassim, N.E., Frail, D.A.,
Goss, W.M., Winkler, P.F., \& Williams, B.F.\ 1997, \aj, 113, 1379
\bibitem{} Green, A. J., Frail, D. A., Goss, W. M., Otrupcek, R.
1997, \aj, 114, 2058
\bibitem{} Harrus, I. M., Hughes, J. P., Singh, K. P., Koyama, K.,
\& Asaoka, I. 1997, \apj, 488, 781
%\bibitem{} Kane, J., Drake, R.P., \& Remington, B.A.\ 1999, \apj, 511, 335
\bibitem{} Koo, B.-C.\ \& Heiles, C.\ 1995, \apj, 442, 679
\bibitem{} Lai, D.\ 2003, in 3D Signatures of Stellar Explosion, a workshop honoring J.C. Wheeler's 60th Birthday (astro-ph/0312542)
\bibitem{} Lockett, P., Gauthier, E., \& Elitzur, M., 1999, \apj, 511, 235
\bibitem{} McKee, C.F.\ \& Cowie, L.L.\ 1975, \apj, 195, 715
%\bibitem{} McKee, C.F.\ \& Ostriker, J.P.\ 1977, \apj, 218, 148
%\bibitem{} Mewe, R., Kaastra, J.S., \& Liedahl, D.A. 1995, Legacy, 6, 16
\bibitem{} Moffett, D. A., \& Reynolds, S. P., 1994, \apj, 425, 668
\bibitem{} Morrison, R., \& McCammon, D., 1983, \apj, 270, 119
\bibitem{} Park, S., Hughes, J.P., Slane, P.O., Burrows, D.N., Warren, J.S.,
 Garmire, G.P., \& Nousek, J.A.\ 2003, \apjl, 592, L41
%\bibitem{} Park, S., Hughes, J.P., Burrows, D.N., Slane, P.O.,
% \& Nousek, J.A.\ 2003b, APJL, astro-ph/0309271
\bibitem{} Pavlov, G.G., Sanwal, D., Kiziltan, B., \& Garmire, G.P.\
2001, \apjl, 559, L131
\bibitem{} Petruk, O.\ 2001, \aap, 371, 267
\bibitem{} Reach, W.T., \& Rho, J.H., 1996, \aap, 315, L277
\bibitem{} ------------------, 1998, \apjl, 507, L93
\bibitem{} ------------------, 1999, \apj, 511, 836
\bibitem{} ------------------, 2000, \apj, 544, 858
\bibitem{} Reach, W.T., Rho, J.H., \& Jarrett, T.H., 2002, \apj, 564, 302
\bibitem{} Reynolds, S.P., \& Moffett, D.A. 1993, \aj, 105, 2226 %(RM93)
\bibitem{} Rho, J.H., \& Petre, R. 1996, \apj, 467, 698 %(RP96)
\bibitem{} Rho, J.H., \& Petre, R. 1998, \apjl, 503, L167
\bibitem{} Rho, J.H., Petre, R., Schlegel, E.M., \& Hester, J.J.\ 1994, \apj, 430, 757
\bibitem{} Shelton, R.  L., Cox, D. P., Maciejewski, W., Smith, R. K.,
Plewa, T., Pawl, A., \& R\'{o}zyczka, M. 1999, \apj, 524, 192
\bibitem{} Shelton, R.  L., Kuntz, K. D., \& Petre, R.\ 2004, \apj, accepted,
astro-ph/0407026
\bibitem{} Slane, P., Smith, R., Hughes, J.P., \& Petre, R.\ 2002, \apj, 
564, 284
\bibitem{} Spitzer, L. Jr. 1962, Physics of Fully Ionized Gases
(New York: Interscience)
%\bibitem{} Strom, R., Johnston, H.M., Verbunt, F., \& Aschenbach, B.\
%1995, Nature, 373, 590
\bibitem{} Turner, T.J.\ \& Pounds, K.A. 1989, \mnras, 240, 833
\bibitem{} Wang, Q.D., Chaves, T., \& Irwin, J.A.\ 2003, \apj, 598, 969
\bibitem{} Wang, Z.R.\ \& Seward, F.D.\ 1984, \apj, 279, 705
%\bibitem{} Warren, J.S., Hughes, J.P., \& Slane, P.O.\ 2003, \apj, 583, 260
\bibitem{} Wardle, M.\ 1999, \apjl, 525, L101
\bibitem{} Wilner, D.J., Reynolds, S.P., \& Moffett, D.A. 1998, \apj, 115, 247
\bibitem{} White, R.L., \& Long, K.S. 1991, \apj, 373, 543 %(WL91)
\bibitem{} Wolszczan, A., Cordes, J.M., \& Dewey, R.J.\ 1991, \apjl, 372, L99
\bibitem{} Xu, J.\ \& Stone, J.M.\ 1995, \apj, 454, 172
\bibitem{} Yusef-Zadeh, F., Wardle, M., Rho, J., \& Sakano, M.\ 2003,
\apj, 585, 319
\bibitem{} Zel'dovich, Ya.\ B. \& Raizer, Yu.\ M.\ 1967, Physics of
shock waves and high-temperature hydrodynamic phenomena
(New York: Academic Press)
\end{thebibliography}
\end{document}